\documentclass[showpacs,prd,amsmath,nofootinbib,amssymb]{revtex4}
\usepackage{graphicx,color}
\usepackage{subfig}
\usepackage{amssymb}
\usepackage{hyperref}
\usepackage[utf8]{inputenc}
\usepackage[english]{babel}
\usepackage{epsfig}
\usepackage{wasysym}
\usepackage{color,xcolor}
\usepackage{pdflscape}
\usepackage{amsmath}
\usepackage{bm}
\usepackage{epsfig}
\usepackage{amsfonts}
\usepackage{dcolumn}
\usepackage{float}

\begin{document}

\title{ Is it  possible to distinguish between different black hole solutions using the Shapiro time delay? \\}

\author{Ednaldo L. B. Junior $^{(1)} $ \footnote{E-mail 
address: ednaldobarrosjr@gmail.com},Manuel E. Rodrigues$^{(2,3)}$\footnote{E-mail 
address: esialg@gmail.com}, Henrique A. Vieira$^{(2)}$\footnote{E-mail 
address: henriquefisica2017@gmail.com}, 
}

\affiliation{$^{(1)}$Faculdade de Engenharia da Computação, Universidade Federal do Pará, \\ Campus Universitário de
Tucuruí, \\ CEP: 68464-000, Tucuruí, Pará, Brazil \\
$^{(2)}$Faculdade de F\'{i}sica, Programa de P\'{o}s-Gradua\c{c}\~{a}o em F\'{i}sica, Universidade Federal do Par\'{a}, 66075-110, Bel\'{e}m, Par\'{a}, Brazil \\
$^{(3)}$Faculdade de Ci\^{e}ncias Exatas e Tecnologia, Universidade Federal do Par\'{a} Campus Universit\'{a}rio de Abaetetuba, 68440-000, Abaetetuba, Par\'{a}, 
Brazil\\}

\begin{abstract}
In this paper we propose to use Shapiro time delay as a tool to distinguish between different black hole solutions. We calculate the analytic Shapiro time, using first-order expansions, for four solutions. They are Schwarzschild, Reissner-Nordstr\"o{}m, Bardeen, and Ayón-Beato and García. We created a numerical experiment, based on measurements made in the solar system, consisting of the round trip path of light passing through a black hole at the center. We obtained different delay times varying between the order of $10^{-4}$s and $10^{-6}$s, for a stellar black hole; 
and variations of the order of hours, for a supermassive one. Considering that the accuracy currently achieved in measurements made in the solar system are on the order of $10^{-12}$s, we believe that this mechanism could be used in black hole model determination in the near future.

\end{abstract}

\date{\today}

\maketitle

\section{INTRODUCTION}

 In 1916, Karl Schwarzschild \cite{SZ} solved Einstein's newly proposed equations of general relativity \cite{Einstein1905} and introduced to the world what we know today as black holes.  These physical entities are defined as a region of spacetime covered by an event horizon from which not even light can escape \cite{Wlad}.  This first exact solution of Einstein's equations, the Schwarzschild black hole, is obtained by considering only mass, there is no angular momentum or electric charge, and therefore can be used as an approximation to describe celestial bodies with low rotation, for example the Sun or even the Earth. Of course it is also possible to get more general exact solutions that consider rotation, electric or magnetic charge, and even  both together \cite{Herlt}. There is an increasing attention in this area  since the 1960s because of the discovery of compact objects and, 
more recently, the first image of the shadow of what is believed to be a supermassive black hole \cite{fotoBN1,fotoBN2,fotoBN3,fotoBN4,fotoBN5,fotoBN6}. However, there is an open problem that still causes discussion among physicists: the singularity. All the solutions mentioned above, for example, have a singularity in origin.

A curvature singularity is understood as a sudden termination point of the geodesics, equations that describe the motion of free particles in general relativity, where quantities such as the density of matter become infinity. There is no final consensus on the subject, but most of the scientific community believes that the singularities present in general relativity are a flaw because it is a classical theory. In this sense, an alternative to the problem is to look for solutions of the Einstein equations that are singularity free. The solution proposed by James Bardeen \cite{Bardeen1} is the first one that has achieved this goal, but it lacked a plausible source of matter. Ayón-Beato and García \cite{Ayon-Beato:1998hmi} proposed, in 1998, the first exact regular solution that had as its source a self-gravitating magnetic charge described by nonlinear electrodynamics \cite{Born}. Later the same authors showed that Bardeen's solution could be described in the same way \cite{Beato}, and it was also realized that one can use nonlinear electrodynamics to construct regular black hole \cite{regular4}. Now we have a wide range of regular Bardeen's type solutions. Some solutions possess electric charge as source \cite{regular2,regular5}, and others  also possess angular momentum \cite{regular6,regular7}.
 Also, the solution was extended to alternative theories of gravity, such as the
$f(r)$ theory \cite{regular1,regular3}, the $f(G)$ theory \cite{regular9,regular8}, and the Rainbow Gravity \cite{regular10}. There are also in the literature works that shows how the regularity of the model is lost when a singular solution is attached to it \cite{Henrique,Rodrigues:2022zph}.

A natural question that arises in this context of black holes is: since these bodies do not emit light, would it be possible to distinguish between different solutions? V. Bozza \cite{Bozza:2002zj}, studying gravitational lenses in the strong field limit,  proposed that the Very Long Baseline Interferometry (VLBI)  should be able to reconstruct the
strong field limit coefficients and select a precise black hole model. In comparing the Bardeen and Reissner-Nordstr\"o{}m solutions, using  absorption \cite{Macedo:2014uga} and  scattering \cite{Macedo:2015qma} of planar massless scalar waves, the authors conclude that the behavior of both is similar, especially when high frequency waves are considered.  Parallel to this result, two papers comparing Reissner-Nordstr\"o{}m and Ayón-Beato and García; considering geodesics analysis and absorption \cite{Paula:2020yfr}, and then scattering \cite{dePaula:2022kzz} of a massive scalar field; proposed that it would be impossible to distinguish regular black holes solutions from standard ones.  In \cite{Kumar:2020yem} 
the conclusion is the same when comparing Kerr's solution with regular ones.
Opposing this interpretation, Z. Stuchlík and J. Schee \cite{Stuchlik:2019uvf} claim that due to nonlinear electrodynamics
we can distinguish between the shadows generated by a Bardeen and Reissner-Nordstr\"o{}m black hole. This results can be reinforced by \cite{Lima:2021las} which shows that, in general, black holes described by different solutions do not generate equal shadows.

In addition to these aforementioned black hole-light interactions; absorption, scattering and shadow formation; there is another interesting effect to be analyzed: the delay of a light signal. This phenomenon is known as the Shapiro time delay  \cite{Shapiro:1964uw} and served as the fourth test of general relativity. The experiment by Shapiro's group sought to measure the time for a light signal sent from Earth to Venus and reflected back to Earth while these planets were at superior conjunction, that is, aligned in a straight line passing through the Sun. The prediction of general relativity is that the curvature caused by mass, in this case the Sun, causes a time delay compared to the flat case. For the Earth-Venus situation this delay should be of the order of \cite{Inverno}

\begin{equation}
    \Delta t \simeq \frac{4 G M_{\odot} }{c^3} \Bigl[ \ln{\left(  \frac{4 D_E D_V}{R_{\odot}^2} \right)} +1  \Bigr],
    \label{eq:one}
\end{equation}
where $G = 6.67408 \times 10^{-11} \text{$Nm^2/kg^2$}$ is the Newton's constant of universal gravitation,  $M_{\odot} = 1.98892 \times 10^{30} \text{kg}$ is the solar mass, $c= 2.99 \times 10^8$ m/s is the speed of light in vacuum , $D_E = 1.495978707 \times 10^{11}  \text{m}$ is radius of Earth's orbit and  $D_V = 1.08 \times 10^{11} $m is radius of Venus orbit both  during superior conjunction, and 
$R_{\odot} = 6.957 \times 10^{8}$ m is the solar radius. These values lead to $\Delta t = 252.282 \mu$s and
the experimentally measured by Shapiro's group was approximately $200 \mu$s.  At the time, the agreement between the theoretical prediction and the experimental value was better than 5\%. Latter, the Viking mission to Mars, in 1976, achieved a better result than 1\% \cite{Will:2014kxa}.  In 2003, the  Cassini spacecraft, in mission to Saturn,   used Shapiro time to measure the PPN $\gamma$ parameter, the prediction of general relativity is that it is unitary,  and found $\gamma = 1 + \left( 2.1 \pm 2.3 \right) \times 10^{-5}$ \cite{Cassini}.

The equation \eqref{eq:one} is obtained by considering the Schwarzschild metric (we will show in more detail how to obtain the Shapiro time for any metric later on) starting from a first order expansion in terms of the mass, and as we commented, this result has a good correspondence with the experimental measurements.  However, we know that almost everything in the universe rotates, so I.G. Dymnikova \cite{Dymnikova} obtained the time delay by considering Kerr's metric. Another way to refine the theoretical prediction is to consider higher order terms in the expansion in terms of the mass \cite{He:2016xiu}, or else to assume a velocity for the black hole \cite{He:2016cya}. Guoxu Feng and Jun Huang \cite{Guoxu} have demonstrated that this effect can be achieved using a purely optical point of view.
In 2001, S. M. Kopeikin \cite{Kopeikin:2001tq}proposed an interesting use for Shapiro time, he
claimed that a measurement of the time delay of light from a quasar
as the light passed by the planet Jupiter could be used to measure the speed of gravity $c_g$.
His line of reasoning was that the velocity $v$ of a body like Jupiter would cause corrections in Shapiro time of the order of $v/c_g$, whereby the speed of gravity could be different from that of light $c_g \neq c$ \cite{Kopeikin:2002tt,Kopeikin:2003kt}.  On 2002, S. M. Kopeikin and  E. B. Fomalont made precise measurements of the
Shapiro delay with $10^{12}$s timing accuracy. They claimed that the  correction term is about 20 percent \cite{Fomalont:2003pd}. Despite achieving high accuracy in time measurement, C. M. Will \cite{Will:2003yj} latter pointed out that this  effect does not depend on the speed of propagation of gravity, but rather only depends on the speed of light. Past the period when it was used as a proof of the theory of general relativity, the Shapiro time began to be useful in monitoring Pulsars \cite{Kaspi,Stairs:1997kz,Ben-Salem:2022txj,Camilo} and in measuring the mass of stars \cite{Fonseca:2021wxt}.

With this context in mind, we will calculate the Shapiro time delay for four solutions; Schwarzschild, Reissner-Nordstr\"o{}m, Bardeen and Ayón-Beato and García. Then, we will propose a idealized experiment that consists in measure a round-trip travel of a light signal sent from a planet to another when they are in superior conjunction, with the central body being a black hole described by one of this solutions. We will study this situation to respond the question: is it  possible to distinguish between different black hole solutions using the Shapiro time delay?
This paper will be organized as follows: in section \ref{sec:one} we will briefly review the main aspects of the four solutions that will be discussed; in section \ref{sec:two} we will calculate the Shapiro time for any metric; in section \ref{sec:three} we will define and calculate the impact parameter, also for any metric; in section \ref{sec:four} we will make the comparison for two black holes with equal event horizons and similar masses; in section \ref{sec:five} we will make the comparison by equalizing both the event horizon and the masses; and in section \ref{sec:six} we will draw our conclusions. We will consider throughout this work the metric signature $(-,+,+,+)$.  Also, we will use, unless otherwise stated, geometrized units where $G =  c =1$.

\section{Black Hole Solutions
\label{sec:one}}
In this section we will show how to obtain the solutions that are used here as examples. Also, we will show some details about it's behavior, such as the event horizon and the regularity of spacetime generated by then. 
\subsection{Schwarzschild 
\label{subsec:Sch}}
The Schwarzschild black hole is a exact solution of the Einstein field equations for vacuum.
It is  a static and spherically symmetric solution,  described by a single parameter, the mass $m$. In the units we are considering, geometrized, it is written as 
\cite{Inverno}
\begin{equation}
    ds^2= - \left( 1- \frac{2m}{r} \right)dt^2 + \left( 1- \frac{2m}{r} \right)^{-1}dr^2 +r^2d \theta^2 +r^2 \sin^2 \theta d \phi^2.
    \label{eq:metricaSC}
\end{equation}
This solution is singular at the origin, that is, one or more components of the Riemann tensor diverges.  A simple way to check this is to analyze the Kretschmann scalar \cite{Lobo:2020ffi}
\begin{equation}
    K = R_{\mu \nu \alpha \beta}R^{\mu \nu \alpha \beta} =  \frac{48m^2}{r^6}.
\end{equation}
From the above result we see that this scalar  diverge when $r \rightarrow 0$. At $r=2m$ the $g_{tt}$ component is zero, and the $g_{rr}$ diverge. This coordinate, once considered a singularity, is today called the event horizon.  The Schwarzschild solution  has only one horizon, while the other solutions we will consider have two.

\subsection{Reissner-Nordstr\"o{}m
\label{subsec:RN}}

\begin{figure}[htpb]
    \centering
    \includegraphics[scale=0.5]{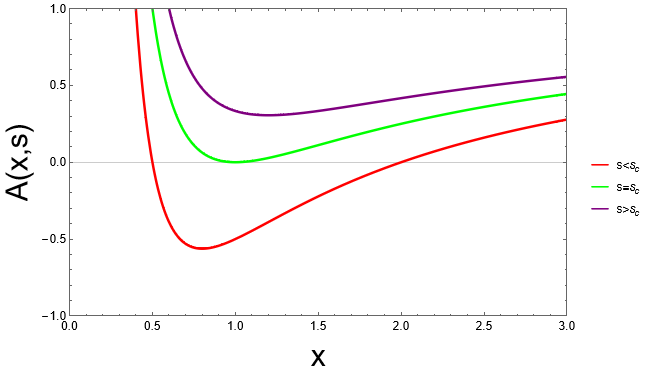}
    \caption{Graphic representation of the auxiliary function $A(x,s)$ for Reissner-Nordstr\"o{}m metric. }
    \label{fig:RN}
\end{figure}

The Reissner-Nordstr\"o{}m metric is a charged, static, and spherically symmetric solution of the Einstein-Maxwell field equations given by 
\begin{equation}
    ds^2= - \left( 1- \frac{2m}{r} +\frac{q_{RN}^2}{r^2} \right)dt^2 + \left( 1- \frac{2m}{r} +\frac{q_{RN}^2}{r^2}  \right)^{-1}dr^2 +r^2d \theta^2 +r^2 \sin^2 \theta d \phi^2.
    \label{eq:metricaRN}
\end{equation}
The parameters that describe it are the electric charge $q_{RN}$ and the ADM mass $m$.
As we said in the introduction, this is a singular solution, we can verify this by analyzing the Kretschmann scalar
\begin{equation}
    K = \frac{4 \left(\frac{2 m}{r}-\frac{q_{RN}^2}{r^2}\right)^2+r^4 \left(\frac{6 q_{RN}^2}{r^4}-\frac{4 m}{r^3}\right)^2+4
   r^2 \left(\frac{2 m}{r^2}-\frac{2 q_{RN}^2}{r^3}\right)^2}{r^4},
\end{equation}
which diverges when $r \rightarrow 0$. This metric has up to two event horizons that can be found by doing $-g_{tt} = 0$, which results in
\begin{equation}
    r_{\pm} = m \pm \sqrt{m^2 -q^2}.
\end{equation}
Note that the event horizons degenerate into only one for a charge, called critical or extreme,  $q_{RN}^{ext} = m $.
Making the following substitutions $x = r/q_{RN}$ and $s= q_{RN}/2m$, and writing the auxiliary function
\begin{equation}
  - g_{tt} = A(x,s) \equiv 1-\frac{1}{s x}+\frac{1}{x^2},
\end{equation}
 Deriving the above expression and equating to zero we find a minimum value $x_m = 2s$. Then, imposing $A(x_m,s) = 0$ we find the critical value  $s_c = 1/2$.  
This auxiliary function carries the same behavior as the metric function, that is, for values of $s<s_c$ we have two event horizons, for $s=s_c$ the horizons degenerate into just one, and when $s>s_c$ there is no event horizon. We  plot of the function $A(x,s)$ in figure \ref{fig:RN}.

\subsection{Bardeen
\label{subsec:Ba}}
\begin{figure}[htpb]
    \centering
    \includegraphics[scale=0.5]{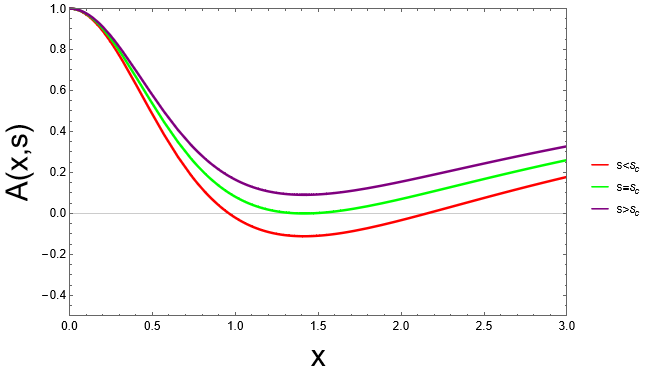}
    \caption{Graphic representation of the auxiliary function $A(x,s)$ for Bardeen metric. }
    \label{fig:um}
\end{figure}

The Bardeen regular solution is a exact, static, and spherically symmetric  solution of the Einstein equations minimally coupled to nonlinear electrodynamics. It is given by
\begin{equation}
    ds^2 = - \left( 1 - \frac{2 M(r)}{r} \right) dt^2 + \left( 1 - \frac{2 M(r)}{r} \right)^{-1} dr^2 +r^2 d \theta^2 + r^2 \sin^2{\theta} d \phi^2,
\end{equation}
with
\begin{equation}
    M(r) = \frac{2 m r^3}{\left(q_{BD}^2+r^2\right)^{3/2}}.
\end{equation}
where $s = |q_{BD}|/2m$, $q_{BD}$ is the magnetic charge, and $m$ is the ADM mass. 
To check for spacetime singularities  we calculate the Kretschmann scalar \cite{Lobo:2020ffi}
\begin{equation}
  K =  \frac{12 m^2 \left(8 q_{BD}^8-4 q_{BD}^6 r^2+47 q_{BD}^4 r^4-12 q_{BD}^2 r^6+4 r^8\right)}{\left(q_{BD}^2+r^2\right)^7},
\end{equation}
which is  regular everywhere. As pointed out in the literature, if we make the the substitutions $x = r/|q_{BD}|$ and $s = |q_{BD}|/2m$, the function $g_{tt}$ can be rewritten as
\begin{equation}
    - g_{tt} = A(x,s) \equiv 1-\frac{x^2}{s \left(x^2+1\right)^{3/2}}.
\end{equation}
Deriving the above expression and equating to zero we find a minimum value $x_m = \sqrt{2}$ independently of the  value of $s$. Then, imposing $A(x_m,s) = 0$ we find the critical value  $s_c = 2/3\sqrt{3}$.  
What happens is that for values of $s<s_c$ we have two event horizons, for $s=s_c$ the horizons degenerate into just one, and when $s>s_c$ there is no event horizon. The function $A(x,s)$ is shown in the figure \ref{fig:um}.
In terms of the magnetic charge and mass the extremization condition is 
\begin{equation}
q_{BD}^{ext} = 2 m s_c = \frac{4}{3 \sqrt{3}} m .
\end{equation}

To summarize, the Bardeen solution, the first regular one, can be obtained by considering a self gravitating magnetic charge as the source. The action used  describes a nonlinear electrodynamics minimally coupled to General Relativity. One can find more information about this black hole in \cite{Beato,Macedo:2014uga,Macedo:2015qma,Zhou:2011aa,Shamir:2020gzh} and references therein.

\subsection{Ayón-Beato and García
\label{subsec:ABG}}
The Ayón-Beato and García regular solution is a static, and spherically symmetric  solution of the Einstein equations minimally coupled to nonlinear electrodynamics, and it is the firs exact regular one. It is given by
\begin{equation}
    ds^2 = - f(r)dt^2 + \left( f(r) \right)^{-1}dr^2 +r^2 d \theta^2 + r^2 \sin^2{\theta} d \phi^2,
\end{equation}
with 
\begin{equation}
    f(r) = 1-\frac{2 m r^2}{\left(q_{ABG}^2+r^2\right)^{3/2}}+\frac{q_{ABG}^2 r^2}{\left(q_{ABG}^2+r^2\right)^2},
    \label{eq:metricaABG}
\end{equation}
where $q_{ABG}$ is the eletric charge and $m$ is the ADM mass. 

\begin{figure}[htpb]
    \centering
    \includegraphics[scale=0.5]{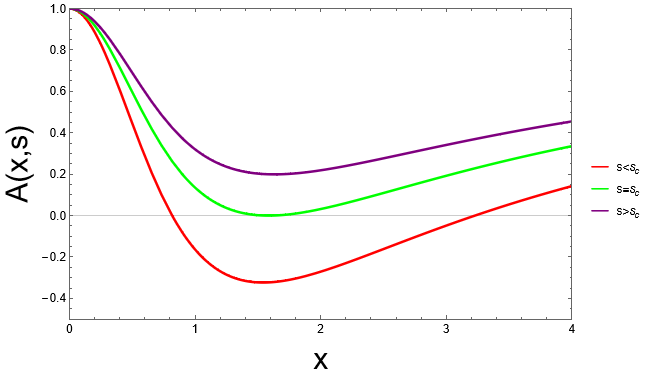}
    \caption{Graphic representation of the auxiliary function $A(x,s)$ for  Ayón-Beato and García metric.}
    \label{fig:dois}
\end{figure}

The Kretschmann scalar for this metric is
\begin{equation}
\begin{aligned}
    &K =  4 \left(\frac{2 m}{\left(q_{ABG}^2+r^2\right)^{3/2}}-\frac{q_{ABG}^2}{\left(q_{ABG}^2+r^2\right)^2}\right)^2+\frac{16 \left(m
   \left(-2 q_{ABG}^4-q_{ABG}^2 r^2+r^4\right)+q_{ABG}^2 (q_{ABG}-r) (q_{ABG}+r)
   \sqrt{q_{ABG}^2+r^2}\right)^2}{\left(q_{ABG}^2+r^2\right)^7} \\ &+\left(\frac{2 q_{ABG}^2 \left(q_{ABG}^4-8 q_{ABG}^2 r^2+3
   r^4\right)}{\left(q_{ABG}^2+r^2\right)^4}-\frac{2 m \left(2 q_{ABG}^4-11 q_{ABG}^2 r^2+2
   r^4\right)}{\left(q^2+r^2\right)^{7/2}}\right)^2,
   \end{aligned}
   \label{eq:krABG}
\end{equation}
which is regular in all spacetime. The function $g_{tt}$  can also be rewritten using the substitutions $x = r/|q_{ABG}|$ and $s = |q_{ABG}|/2m$
\begin{equation}
   - g_{tt}= f(r) = A (x,s) \equiv 1-\frac{x^2}{s \left(x^2+1\right)^{3/2}}+\frac{x^2}{\left(x^2+1\right)^2},
\end{equation}
which has only one real value $x_c$ and $s_c$ as showed in figure \ref{fig:dois}, this values can be found by doing
\begin{equation}
    A(x_c,s_c) = 0, \ \ \ \ \ \frac{\partial A(x_c,s_c)}{\partial x} =0,
\end{equation}
and the result is $x_c \approx 1.58$ and $s_c \approx 0.317$. The interpretation of these critical values is: for $s>s_c$ there is no event horizon,  for $s=s_c$ there is only one degenerated even horizon, and for $s<s_c$ there are two event horizons givem by 
\begin{equation}
    r_{\pm} = |q_{ABG}| \left( \left( \frac{1}{4s} + \frac{\sqrt{u(s)}}{12s} \pm \frac{\sqrt{6}}{12s} \left( \frac{9}{2} -12s^2 -\frac{u(s)}{6} - \frac{9(12s^2-1)}{\sqrt{u(s)}} \right)^{1/2}  \right)^2  -1\right)^{1/2},
\end{equation}
where
\begin{equation}
    u(s) = 6 \left(   \frac{3}{2} - 4s^2 + sv(s)^{1/3} - \frac{4s \left( 11s^2-3 \right)}{v(s)^{1/3}}  \right),
\end{equation}
and
\begin{equation}
    v(s) = 4 \left( 9s +74s^3 + \sqrt{27(400s^6 -112s^4+47s^2-4)}  \right).
\end{equation}
In terms of electric charge and mass we have $q_{ABG}^{ext} \approx 0.634 m$.

So, in a few words, the Ayón-Beato and García black hole is the first exact regular solution of the Einstein equations as one can see in \eqref{eq:krABG}. It is obtained using the non linear electrodynamics minimally coupled to general relativity  and has as its source an electric field. To obtain more explanations about this solution one can look in \cite{Ayon-Beato:1998hmi,Paula:2020yfr,dePaula:2022kzz} and references therein. 

\section{Shapiro Time for a general metric
\label{sec:two}}

 The equations of motion for free particles in general relativity are the so-called geodesic equations. They are the generalization of the straight line to curved spacetime. 
 In order to  obtain the Shapiro time delay we must first write this equation for photons. The null radial geodesic equation  is  \cite{Inverno} 
\begin{equation}
g_{\mu\nu}\dot{x}^{\mu}\dot{x}^{\nu}=0,
\end{equation}
for  a general static and spherically symmetric spacetime we have
\begin{equation}
g_{tt}\dot{t}^2+g_{rr}\dot{r}^2+g_{\theta \theta}\dot{\theta}^2+g_{\phi \phi}\dot{\phi}^2=0,
\label{eq:geogeral}
\end{equation}

where the dot means a differentiation with respect to the affine parameter. Considering that the geodesic is located in equatorial plane $\theta= \pi /2$ and the specific static and spherically symmetric configuration given by 
\begin{equation}
    ds^2 = - B(r) dt^2 +A(r)dr^2 + C(r) \left( d\theta^2 + \sin^2 \theta d\phi^2  \right),
    \label{eq:dsshapiro}
\end{equation}
 the path of a light ray is described by \cite{Bhadra:2008gt,Deng:2017umx}
\begin{equation}
    - B(r) \dot{t^2} + A(r) \dot{r}^2 + C(r) \dot{\phi}^2 = 0.
    \label{eq:geo}
\end{equation}
This movement has two conserved quantities: the total energy, and the angular momentum. They are \cite{Weinberg}
\begin{equation}
    E = B(r) \dot{t},
    \label{eq:energia}
\end{equation}
\begin{equation}
    L = C(r) \dot{\phi}.
    \label{eq:momento}
\end{equation}
With equations \eqref{eq:geo}, \eqref{eq:energia} and \eqref{eq:momento} we can write
\begin{equation}
    \frac{d \phi}{d r} =  \frac{1}{C(r)} \Biggl[   \frac{1}{A(r)B(r)} \left( \frac{1}{b^2} - \frac{B(r)}{C(r)} \right) \Biggr]^{-1/2},
    \label{eq:phi}
\end{equation}
where $b \equiv L / E$ is the impact parameter, we will  detail this quantity in the next section. When  the light reaches the closest approach to the black hole $r=d$ we have $dr/d \phi = 0$, so 
\begin{equation}
    b = \sqrt{\frac{C(d)}{B(d)}}.
    \label{eq:bemd}
\end{equation}
Collecting \eqref{eq:energia}, \eqref{eq:momento}, \eqref{eq:phi} and \eqref{eq:bemd} into the geodesic equation \eqref{eq:geo} we get
\begin{equation}
    \frac{dt}{dr} =  \frac{1}{b} \sqrt{\frac{A(r)}{B(r)}} \Biggl[ \frac{1}{b^2} - \frac{B(r)}{C(r)} \Biggr]^{-1/2}.
    \label{eq:dtdr}
\end{equation}
Then, the total time between the closest approach $d$ and a point $r_1$ is
\begin{equation}
    t(r_1,d) = \frac{1}{b} \int_{d}^{r_1}   \sqrt{\frac{A(r)}{B(r)}} \Biggl[ \frac{1}{b^2} - \frac{B(r)}{C(r)} \Biggr]^{-1/2}dr,
    \label{eq:tintegrado}
\end{equation}
from now on we need to choose specific functions for $A(r),B(r)$ and $C(r)$. Let's consider the Schwarzschild solution, which is a  good approximation for the solar system. Equation \eqref{eq:dsshapiro} for this solution is
\begin{equation}
    ds^2 = - \left( 1- \frac{2m}{r} \right) dt^2 +\left( 1- \frac{2m}{r} \right)^{-1}dr^2 + r^2 \left( d\theta^2 + \sin^2 \theta d\phi^2  \right),
\end{equation}
then \eqref{eq:tintegrado} is
\begin{equation}
      t(r_1,d) = \int_{d}^{r_1}  \frac{r^2\sqrt{\left(d-2m \right)r}dr}{(r-2 m) \sqrt{ 2 d^3 m-d^3 r+r^3 (d-2 m)}},
\end{equation}
 although the above integral can be solved analytically \cite{Valeri}, we will use a Taylor series expansion and so the total time is 
\begin{equation}
    t(r_1,d) = \sqrt{r_1^2 -d^2} + m \left(  \frac{r_1-d}{r_1+d} \right)^{1/2} + 2m \ln{\left( \frac{r_1 + \sqrt{r_1^2-d^2}}{d} \right)} + \mathcal{O} (m^2).
    \label{eq:ttotalSC}
\end{equation}
 Note that the first term of \eqref{eq:ttotalSC} is  the result for flat spacetime.  The time to a light signal go from a point $r_1$ to another one $r_2$ and go back for $r_1$ is
 \begin{equation}
     T_{SC} = 2 \left( t(r_1,d) + t(r_2,d) \right)
 \end{equation}
 i.e.
\begin{equation}
\begin{aligned}
  & T_{SC} = 2 \left( \sqrt{r_1^2 -d^2} +\sqrt{r_2^2 -d^2} \right) + 2m \Biggl[ \left( \frac{r_1-d}{r_1+d} \right)^{1/2} + \left( \frac{r_2-d}{r_1+d} \right)^{1/2} \Biggr]
   \\ &+ 4m \ln{ \Biggl[ \left( \frac{r_1 + \sqrt{r_1^2-d^2}}{d} \right) \left( \frac{r_2 + \sqrt{r_2^2-d^2}}{d} \right) \Biggr] }.
   \end{aligned}
   \label{eq:TSC}
\end{equation}
Considering that $r_1>r_2 \gg d$,  the time delay  (just subtract the part related to the flat case from the total) is
\begin{equation}
    \Delta T_{SC} \simeq  4m \Bigl[ \ln{\left(  \frac{4 r_1 r_2}{d^2} \right)} +1  \Bigr],
    \label{eq:delaySC}
\end{equation}
which in international system of units  is exactly \eqref{eq:one}.

\section{Impact Parameter
\label{sec:three}}
In the previous section, we used in our calculations the quantity $b \equiv L/E$ which is the impact parameter. In this section we will show how to obtain it for general metric. But first, it is important to clarify the need to obtain this parameter. Use the same procedure, i.e. do  $d = R_{\odot}$ as we did for the Sun, is impossible for the case of black holes. One could think ingenuously and do for this case $d= r_+$. However, as we will show in a moment, the light (or even particles) is absorbed if it is sent from a distance smaller than the critical impact parameter $b_c$, which is always larger than the event horizon. Since Shapiro's experiment consists of light being only slightly deflected by the massive body, we will need to know $b_c$ of each black hole in order to choose $d$ appropriately.

\begin{figure}[htpb]
    \centering
    \includegraphics[scale=0.5]{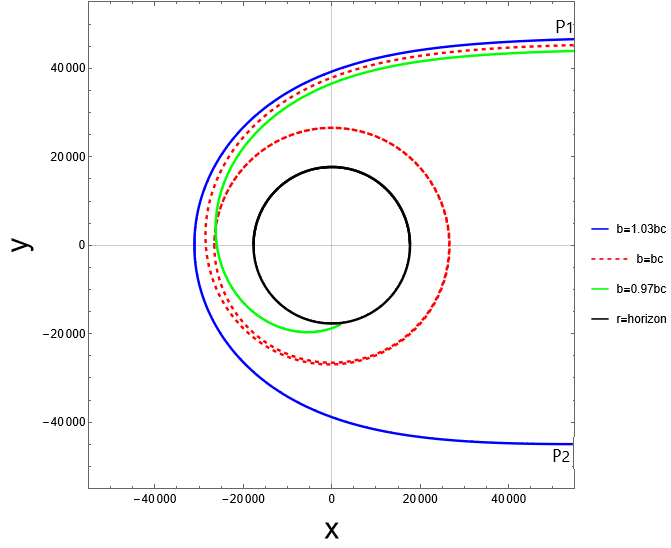}
    \caption{Graphical representation of the light path, considering the Schwarzschild black hole,  for different values of the impact parameter.}
    \label{fig:tres}
\end{figure}

To start, we choose the equatorial plane where $\theta= \pi / 2$, we have from \eqref{eq:geogeral} 
\begin{equation}
g_{tt}\dot{t}^2+g_{rr}\dot{r}^2+g_{\phi\phi}\dot{\phi}^2=0,
\label{eq:geo1}
\end{equation}
we also have the conserved quantities, energy and momentum, respectively given by
\begin{equation}
g_{tt}\dot{t}=E, 
\end{equation}
\begin{equation}
     g_{\phi\phi}\dot{\phi}=L.
\end{equation}
Substituting these quantities into  \eqref{eq:geo1} we have
\begin{equation}
   \dot{r}^2=  E^2-V_{eff}(r),
\end{equation}
where $V_{eff}(r)=(-g_{tt}/r^2)L^2$. Now, we will look for unstable photon orbits given by the conditions
\begin{equation}
    \frac{d V_{eff}(r_c) }{dr} =0,
    \label{eq:con1}
\end{equation}
and
\begin{equation}
    \frac{d^2V_{eff}(r_c)}{dr^2}<0.
    \label{eq:con2}
\end{equation}
In the above equations $r_c$ is called capture radius, this nomenclature will be justified soon after. The first one leads to 
\begin{equation}
    r_c g^{\prime}_{tt} - 2g_{tt} =0,
\end{equation}
where the prime denotes a derivation with respect to the radial coordinate. And the second one is
\begin{equation}
    r_c g^{\prime \prime}_{tt} - g^{\prime}_{tt} < 0. 
\end{equation}
For Schwarzschild solution \eqref{eq:metricaSC}, for example,  we have 
\begin{equation}
    V_{eff}(r) = \frac{L^2 \left(1-\frac{2 m}{r}\right)}{r^2}
\end{equation}
and by applying  \eqref{eq:con1} we get $r_c = 3m$ which leads to 
\begin{equation}
    \frac{d^2V_{eff}(r_c)}{dr^2} = -\frac{2 L^2}{81 m^4},
\end{equation}
since $L$ and $m$ are positive quantities, \eqref{eq:con2} is satisfied. After finding the capture radius, we choose $E = V_{eff}(r_c)$  and get
\begin{equation}
    \dot{\phi}(r_c)= E \left(\frac{-1}{g_{tt}g_{\phi \phi}} \right)^{1/2},
\end{equation}
being the definition of the impact parameter $b \equiv L/E$, we have that it's value considering $r=r_c$ is
\begin{equation}
    b_c = \frac{L(r_c)}{E(r_c)} = g_{\phi \phi} \left(  \frac{-1}{g_{tt}g_{\phi \phi}} \right)^{1/2} \arrowvert_{r=r_c}.
\end{equation}
For Schwarzschild solution $b_c = 3 \sqrt{3} m$. We know that the radius of the event horizon for this solution is $r=2m$, then the capture radius $r_c$ and the critical impact parameter $b_c$ are larger. In figure \ref{fig:tres} we illustrate the situation of light rays being sent from a point $P_1$ with different impact parameters.  The black line corresponds to the event horizon of  a Schwarzschild black hole.
Note that for $b=b_c$, red dotted line, the photons are captured and start orbiting the black hole with radius $r=r_c$ (the capture radius), in this case it is an unstable orbit. When light is sent with $b< b_c$ it is absorbed by the black hole, green line.  But, when we consider $b > b_c$, blue line, the trajectory is only slightly deviated and the rays are able to escape the gravity of the black hole and reach the point $P_2$.

\section{Comparison between two black holes with the same event horizon and similar masses
\label{sec:four}}
In this section we will begin to compare the Shapiro time for different black hole solutions. It is worth remembering that our goal is to check if is it possible
 to distinguish two black holes using this feature.
To begin with, we will compare the solutions of  Schwarzschild and Reissner-Nordstr\"o{}m. 

It would be interesting if we could make the event horizons and masses equal. However, if we impose that the horizons are equal and so write the equation
\begin{equation}
    2 m_1 = m_2 + \sqrt{m_2^2 - q_{RN}^2},
    \label{eq:horizons1}
\end{equation}
where $m_1$ and $m_2$ are the masses, respectively, of the black holes of Schwarzschild and Reissner-Nordstr\"o{}m. Solving \eqref{eq:horizons1} we find that the charge $q_{RN}$ is
\begin{equation}
    q_{RN} = 2 \sqrt{m_1 m_2 - m_1^2},
    \label{eq:carga1}
\end{equation}
this result implies that $m_1<m_2 < 2m_1$. Therefore the comparison will be considering close masses. 

The situation considered is: what is the time delay of a light signal, traveling back and forth, from a planet $P_1$, with radius $r_1$, to another planet $P_2$, with radius $r_2$. We will admit that the planets are in superior conjunction, and $r_1>r_2 \gg d$, with $d$ being the distance of closest approach to the black hole that is positioned between the planets. We will always use  $d$ as being greater than the critical impact parameter $b_c$ of the solutions considered, so we will deal with a situation similar to the one shown in the figure \ref{fig:tres}. We stress that this model is far idealized compared to those used to measure the Shapiro effect in the solar system, but it is a good approximation to answer our motivating question.
We have already calculated the total travel time \eqref{eq:TSC} and the delay \eqref{eq:delaySC} for the Schwarzschild balck hole. 
Before trying to calculate the integral \eqref{eq:tintegrado}  for metric \eqref{eq:metricaRN} we need to make a small adjustment  because, as it stands, the Taylor series expansion in terms of mass yields an inconsistent result.  We expect that the time found for  Reissner-Nordstr\"o{}m solution is the same as \eqref{eq:TSC} at the limit $q_{RN} \rightarrow 0$, but we do not find that  using the mass expansion for solve the integral \eqref{eq:tintegrado}. So, we will use the metric as
\begin{equation}
\begin{aligned}
   & B(r)  = 1 -\frac{2 G m_1}{r}+\frac{G q_{RN}^2}{r^2}, \\
   & A(r)  = B(r)^{-1}, \\
   & C(r) = r^2.
    \end{aligned}
    \label{eq:metricaRN2}
\end{equation}
The justification for this insight is that, in the S.I, we have 
\begin{equation}
    -g_{tt} = 1 - \frac{2G m_1 }{c^2r}
\end{equation}
for the Schwarzschild, and in this case, the expansion  mentioned in section \ref{sec:two} is the same if we consider $m_1$ or $G$. Substituting \eqref{eq:metricaRN2} in \eqref{eq:tintegrado} we get
\begin{equation}
    t(r_1,d) = \int_{d}^{r_1}  \frac{r^4}{G \left(q_{RN}^2-2 m_2 r\right)+r^2} \left( \frac{d^2 r^4-\left(d^4 \left(G \left(q_{RN}^2-2 m_2
   r\right)+r^2\right)\right)-2 d G m_2 r^4+G q_{RN}^2 r^4}{d^2-2 d G m_2+G q_{RN}^2} \right)^{-1/2} dr,
   \label{eq:integralRN}
\end{equation}
and, after a Taylor series expansion, we can integrate the above equation to encounter
\begin{equation}
     t(r_1,d) = \sqrt{r_1^2 -d^2} -\frac{3 G q_{RN}^2 \cot ^{-1}\left(\frac{d}{\sqrt{r_1^2-d^2}}\right)}{2 d}
     +G m_2 \left( \frac{ r_1-d}{r_1+d}   \right)^{1/2}      +4 G m_2 \tanh
   ^{-1}\left(\sqrt{\frac{r_1-d}{r_1+d}}\right)   + \mathcal{O} (G^2).
   \label{eq:ttoltalRN}
\end{equation}
Note that if we put $q_{RN}=0$ and $G=1$ in \eqref{eq:ttoltalRN} the result will be exactly \eqref{eq:ttotalSC} \cite{Integrais}. 
Note that 
 \begin{equation}
   tanh^{-1} (u) = \frac{1}{2} ln \left( \frac{1+u}{1-u} \right).
\end{equation}
The charge part in the above equation has the negative sign, so it will decrease the difference between the total relativistic time and the time for the plane case, this effect is opposite to that of mass. 
The total travel time is
\begin{equation}
    \begin{aligned}
       & T_{RN} = 2 \left(t(r_1,d) + t(r_2,d) \right) = 2 \left( \sqrt{r_1^2 -d^2} + \sqrt{r_2^2 -d^2} \right) -\frac{3 G q_{RN}^2 \biggl[ \cot ^{-1}\left(\frac{d}{\sqrt{r_1^2-d^2}}\right) + 
        \cot ^{-1}\left(\frac{d}{\sqrt{r_2^2-d^2}}\right)  \biggr]}{ d} \\ & +2G m_2 \Biggl[  \left( \frac{ r_1-d}{r_1+d}   \right)^{1/2}  +  \left( \frac{ r_2-d}{r_2+d}   \right)^{1/2}  \Biggr] 
        +8 G m_2 \Biggl[ \tanh^{-1}\left(\sqrt{\frac{r_1-d}{r_1+d}}\right) + \tanh^{-1}\left(\sqrt{\frac{r_2-d}{r_2+d}}\right) \Biggr],
    \end{aligned}
\end{equation}
and the Shapiro time delay, with $r_1 \gg d$ and $r_2 \gg d$ is,
\begin{equation}
    \Delta T_{RN} = 4 G m_2 \Biggl[ \ln\left(\frac{4 r_1 r_2}{d^2}\right)+1 \Biggr] + 3Gq_{RN}^2 \left(\frac{1}{r_1} + \frac{1}{r_2} - \frac{\pi }{d} \right).
   \label{eq:delayRN}
\end{equation}
 
Looking at \eqref{eq:ttoltalRN}, \eqref{eq:delayRN} expressions and those referring to Schwarzschild solution \eqref{eq:TSC} and \eqref{eq:delaySC} we could try to conclude that it is in fact possible to distinguish between them through round-trip light signal experiments. But this would be a hasty conclusion, because such measurements have a certain limit of accuracy that cannot always meet the demands of the theoretical model. 

The last step before proceeding to the numerical comparison is to calculate the critical impact parameter for the Reissner-Nordstr\"o{}m solution. Following the procedure described in section \ref{sec:three} one finds 
\begin{equation}
    b_c^{RN} = \frac{1}{2} \left(\sqrt{9 m_2^2-8 q_{RN}^2}+3 m_2\right)^2 \sqrt{\frac{1}{2 m_2 \left(\sqrt{9
   m_2^2-8 q_{RN}^2}+3 m_2\right)-4 q_{RN}^2}}.
\end{equation}

Now, let's begin to compare the Shapiro time delay  for this two solutions. Since we will make a numerical comparison we will choose the values of some constants. We chose the masses as
\begin{equation}
    \begin{aligned}
        &m_1 = 6.0 \times M_{\odot}^{\prime}, \\
        & m_2 = 6.1 \times M_{\odot}^{\prime},
        \label{eq:massas1}
    \end{aligned}
\end{equation}
where $M_{\odot}^{\prime}$ is the solar mass  in the geometrized units 
\begin{equation}
    M_{\odot}^{\prime} = \frac{M_{\odot}}{c^2G^{-1}} = \frac{1.98892 \times 10^{30} \text{kg}}{\left( 2.99 \times 10^{8} \text{m/s} \right)^2 \left( 6.67408 \times 10^{-11} \text{Nm}^2/\text{kg}^2 \right)^{-1}} = 1484.06 \text{m}.
\end{equation}
We reinforce that geometrized units are powers of length (m). The masses in \eqref{eq:massas1} are characteristic of so-called stellar black holes, they lead, by inserting then into \eqref{eq:carga1}, to the charge $q_{RN} = 2299.1$. To wit, these choices lead to the following event horizons
\begin{equation}
    \begin{aligned}
        & 2 m_1 = 17808.8, \\
        & m_2 + \sqrt{m_2^2 -q_{RN}^2} = 17808.8.
    \end{aligned}
\end{equation}
Also, the critical impact parameter for both solutions are
\begin{equation}
    \begin{aligned}
        & b_c^{SC} = 46268.6 ,\\
        & b_c^{RN} = 46527.5.
    \end{aligned}
\end{equation}
Given these values, we will choose $ d = 46600 $. This particular value has nothing special compared to any other greater than $46527.5$, which is the value of the largest critical impact parameter considered. The distances between the planets and $d$ will be
\begin{equation}
    \begin{aligned}
        &r_1 = 2.0 \times 10^{4} d, \\
        &r_2 = 1.0 \times 10^{4} d.
    \end{aligned}
\end{equation}
 In this case, we are respecting the relations $r_1>r_2 \gg d$. These distances are on the order of $10^8$ meters, making the travel time between them on the scale of seconds. 
There is no realistic scenario associated with these choices, but they allow us to use \eqref{eq:delaySC} and \eqref{eq:delayRN}. Then, we have 
\begin{equation}
    \Delta T_{SC} = 765782,
\end{equation}
and
\begin{equation}
    \Delta T_{RN}= 778010.
\end{equation}
We can already tell that there is a difference in the outcome of the solutions, but interpretation using the geometrized units is not our focus. To convert the time to the international system of units (S.I) one just divide the value in geometric units by $c$ (speed of light in  vacuum). So, in S.I we have
\begin{equation}
    \Delta T_{SC}^{\prime} = \frac{\Delta T_{SC}}{c} = 2.55261 \times 10^{-3} s,
\end{equation}
and
\begin{equation}
    \Delta T_{RN}^{\prime}= \frac{\Delta T_{RN}}{c} =2.59337 \times 10^{-3} s .
\end{equation}

Note that the difference in Shapiro  time delay for these solutions is of the order of $10^{-5}$s. Our optical  clocks on ground have sufficient precision to measure this  timescale \cite{Chou,Chou2,Zhang}. Therefore, in this idealized situation, the solutions are distinguishable by the Shapiro delay time. But, of course, in a realistic situation this experiment would require a more precise model, and this timescale could be difficult to reach.  Instead of  chose arbitrary distances and masses one can use the values from the Earth-Sun-Venus case. But note that this would not lead to a better, or even more realistic result than the one presented above. A difference between the delay times would also be detected, only these would be smaller than those we found here.
There is a possibility  to make the difference between times greater than those found in the previous example.
Looking at the expressions  \eqref{eq:delaySC} and \eqref{eq:delayRN} we realize that the delay is proportional to the mass, so let's check how the Shapiro time stand considering a supermassive black hole. For this purpose we will choose now
\begin{equation}
    \begin{aligned}
        &m_1 = 3.4 \times 10^9 \times M_{\odot}^{\prime}, \\
        & m_2 = 3.6 \times 10^9 \times M_{\odot}^{\prime},
        \label{eq:massas2}
    \end{aligned}
\end{equation}
then we have $q_{RN} = 2.44758 \times 10^{12}$, and 
\begin{equation}
      \begin{aligned}
        & b_c^{SC} = 2.62189 \times 10^{13} ,\\
        & b_c^{RN} = 2.67467 \times 10^{13}.
    \end{aligned}
\end{equation}
Considering this values we choose $d = 2.7 \times 10^{13}$ (again, we have freedom of choice for this distance as long as the no-return condition is respected $d>b_c$), 
$r_1 = 2.0 \times 10^{2}d$, and $r_2 = 1.0 \times 10^{2}d$. Here in this example the closest approach distance is of the order of $10^{13}$ meters, that is, two orders of magnitude greater than the distances involved in the Earth-Sun-Venus experiment. For this reason, we chose  $r_1$ and $r_2$ so that distances can be traveled by light on a time scale of years  ($10^{15}$ meters lead to a time travel of approximately $1.05$ years). Given that these parameters are the same for both types of black hole, our object of analysis will obviously not depend on this choice. However, we know that in a real observation situation we are limited to the human time scale. Moreover, we also know that observations from very distant regions (outside the solar system) have more complications and experimental errors compared to those made at shorter ranges.
With this we have that the time delays are
\begin{equation}
    \Delta T_{SC}^{\prime} = 826827 s \approx 9.6 \ \text{days},
\end{equation}
and
\begin{equation}
    \Delta T_{RN}^{\prime}= 872012 s \approx 10.1 \ \text{days} .
\end{equation}
An important detail is that for the two cases considered the delay of the Reissner-Nordstr\"o{}m black hole was greater. This was expected since we consider $m_2 > q_{RN}$ for both examples and this implies that the mass term dominates the equation \eqref{eq:delayRN}. As we can see, the difference in time delay is now in the order of hours.


\section{Comparison between three black holes with the same event horizon and masses
\label{sec:five}}
In this section we will compare the Shapiro time for the solutions of Reissner-Nordstr\"o{}m, Bardeen and Ayón-Beato and García (ABG). We already have the necessary expressions for the first. For the Bardeen metric  the equation \eqref{eq:tintegrado} becomes

\begin{equation}
    t(r_1,d) =   \int_{d}^{r_1}\left( \frac{  \frac{1}{d^2}-\frac{2 m_3}{\left(d^2+q_{BD}^2\right)^{3/2}}}{2 m_3
   \left(\frac{1}{\left(q_{BD}^2+r^2\right)^{3/2}}-\frac{1}{\left(d^2+q_{BD}^2\right)^{3/2}}\right)+\frac{1}{d^2}-\frac{1}{r^2}} \right)^{1/2} \left(1-\frac{2 m_3
   r^2}{\left(q_{BD}^2+r^2\right)^{3/2}}\right)^{-1} dr,
\end{equation}
where $m_3$ is the mass, and $q_{BD}$ is the magnetic charge. Again, we will use a Taylor series expansion to solve the integral, which results in

\begin{equation}
   \begin{aligned}
    &t(r_1,d) = \sqrt{r_1^2-d^2}  -\frac{ m_{3} q_{BD}^2 \left( 3 d^2 +2 q_{BD}^2\right)\sqrt{r_1^2-d^2}}{\left(d^2+q_{BD}^2\right)^2
   \sqrt{r_1^2+q_{BD}^2}}+2 m_{3} \ln\left(\sqrt{r_1^2-d^2}+\sqrt{r_1^2+q_{BD}^2}\right)  -2 m_3 \ln \left(d^2+q_{BD}^2\right) \\
  &+\frac{d^4 m_3 \sqrt{r_1^2+q_{BD}^2}}{\sqrt{r_1^2-d^2}
   \left(d^2+q_{BD}^2\right)^2}-\frac{d^4 m_3}{\sqrt{r_1^2-d^2}
   \left(d^2+q_{BD}^2\right)^{3/2}} + \mathcal{O}(m^2).
   \end{aligned}
   \label{eq:ttotalBD}
\end{equation}

This is the time for the light to go from the  point  of closest approach $d$ to general point $r_1$. Considering the same situation as in the previous section; that is, light being sent from a Planet at $r_1$ to another at $r_2$, these bodies being at superior conjunction, and reflected back to $r_1$; the total time is $2 \left(t(r_1,d) + t(r_2,d) \right)$. Then we have

\begin{equation}
    \begin{aligned}
       & T_{BD}   =  2 \left( \sqrt{r_1^2-d^2} + \sqrt{r_2^2-d^2} \right)    -\frac{2m_{3} q_{BD}^2 \left(3 d^2 +2q_{BD}^2 \right)}{\left(d^2+q_{BD}^2\right)^2} \left( \frac{\sqrt{r_1^2-d^2}}{\sqrt{r_1^2+q_{BD}^2} } + \frac{\sqrt{r_2^2-d^2}}{\sqrt{r_2^2+q_{BD}^2} }\right)   -4 m_3 \ln \left(d^2+q_{BD}^2\right)   \\
     &     +4 m_{3} \ln\Biggl[ \left( \sqrt{r_1^2-d^2}+\sqrt{r_1^2+q_{BD}^2} \right)\left( \sqrt{r_2^2-d^2}+\sqrt{r_2^2+q_{BD}^2} \right) \Biggr]
  +\frac{2d^4 m_3 }{
   \left(d^2+q_{BD}^2\right)^2}  \left( \frac{\sqrt{r_1^2+q_{BD}^2}}{\sqrt{r_1^2-d^2}} +\frac{\sqrt{r_2^2+q_{BD}^2}}{\sqrt{r_2^2-d^2}}  \right) \\ 
   &-\frac{2d^4 m_3}{ \left(d^2+q_{BD}^2\right)^{3/2}} \left( \frac{1}{\sqrt{r_1^2-d^2}} +\frac{1}{\sqrt{r_2^2-d^2}} \right).
    \end{aligned}
\end{equation}
The effect in focus here is only the relativistic travel time delay compared to the flat case,  which is
\begin{equation}
    \begin{aligned}
       & \Delta T_{BD} = 2 \Biggl[-2  m_3 \ln \left(d^2+ q_{BD}^2\right)-2m_3 \left( \frac{1}{\sqrt{\frac{
   q_{BD}^2}{r_1^2}+1}} + \frac{1}{\sqrt{\frac{
   q_{BD}^2}{r_2^2}+1}} \right)    \\
  &  +2 m_3 \ln \left( \left(r_1 \sqrt{\frac{
   q_{BD}^2}{r_1^2}+1}+r_1\right)\left(r_2 \sqrt{\frac{
   q_{BD}^2}{r_2^2}+1}+r_2\right)   \right)  \Biggr],
    \end{aligned}
    \label{eq:delayBD}
\end{equation}
where we use the approximation $r_1 \gg d$ and $r_2 \gg d$. It is worth mentioning that both \eqref{eq:ttotalBD} and \eqref{eq:delayBD} return, as expected, the same results as  Schwarzschild case when the magnetic charge is zero.

For ABG solution \eqref{eq:metricaABG} we have the same integrand as \eqref{eq:integralRN},  this of course making the same consideration of using the Taylor series expansion for the gravitation constant $G$ instead of the mass of the black hole. So, the result is
\begin{equation}
     t(r_1,d) = \sqrt{r_1^2 -d^2} -\frac{3 G q_{ABG}^2 \cot ^{-1}\left(\frac{d}{\sqrt{r_1^2-d^2}}\right)}{2 d}
     +G m_4 \left( \frac{ r_1-d}{r_1+d}   \right)^{1/2}      +4 G m_4 \tanh
   ^{-1}\left(\sqrt{\frac{r_1-d}{r_1+d}}\right)   + \mathcal{O} (G^2).
   \label{eq:ttoltalABG}
\end{equation}
where $m_4$ is the mass, and $q_{ABG}$ is the electric charge.  Taking into account the same situation as mentioned above, the total time is

\begin{equation}
    \begin{aligned}
       & T_{ABG} = 2 \left( \sqrt{r_1^2 -d^2} + \sqrt{r_2^2 -d^2} \right) -\frac{3 G q_{ABG}^2 \biggl[ \cot ^{-1}\left(\frac{d}{\sqrt{r_1^2-d^2}}\right) + 
        \cot ^{-1}\left(\frac{d}{\sqrt{r_2^2-d^2}}\right)  \biggr]}{ d} \\ & +2G m_4 \Biggl[  \left( \frac{ r_1-d}{r_1+d}   \right)^{1/2}  +  \left( \frac{ r_2-d}{r_2+d}   \right)^{1/2}  \Biggr] 
        +8 G m_4 \Biggl[ \tanh^{-1}\left(\sqrt{\frac{r_1-d}{r_1+d}}\right) + \tanh^{-1}\left(\sqrt{\frac{r_2-d}{r_2+d}}\right) \Biggr],
    \end{aligned}
\end{equation}
and the Shapiro time delay, with $r_1 \gg d$ and $r_2 \gg d$, is
\begin{equation}
    \Delta T_{ABG} = 4 G m_4 \Biggl[ \ln \left(\frac{4 r_1 r_2}{d^2}\right)+1 \Biggr]  + 3 G q_{ABG}^2 \left( \frac{1}{r_1} + \frac{1}{r_2} - \frac{\pi}{d} \right) .
   \label{eq:delayABG}
\end{equation}

Note that the Shapiro time  for  Reissner-Nordstr\"o{}m and ABG would only be equal if we consider a situation where $m_2=m_4$ and $q_{RN} = q_{ABG}$. As we will choose to equalize the masses and event horizons we will have different charges, and so the times will also be different.

To begin with, we will choose the masses as
\begin{equation}
    m_2 = m_3 = m_4 = 6.1 \times M_{\odot}^{\prime},
\end{equation}
and the  event horizon for for all three solutions
\begin{equation}
    r_+ = 17808.8.
\end{equation}
These choices immediately lead to
\begin{equation}
q_{RN} =  2299.1.
\end{equation}
We can calculate numerically the other charges $q_{BD}$ and $q_{ABG}$ 
\begin{equation}
   q_{BD} = 1874.62,  \ \ \ q_{ABG} = 1454.32.
\end{equation}
The critical impact parameters are
\begin{equation}
    \begin{aligned}
        & b_c^{RN} = 46527.5, \\
        & b_c^{BD} = 42464.2, \\
        & b_c^{ABG} = 42538.7. 
    \end{aligned}
\end{equation}
By the same intuition as in the previous section, we will choose $d=46600$, since it is larger than these three values. The radii of the planets will be
\begin{equation}
    \begin{aligned}
        & r_1 = 2.0 \times 10^4d, \\
        & r_2 = 1.0 \times 10^4 d.
    \end{aligned}
\end{equation}

We compiled the results obtained through these choices in table \ref{tabela1}. Recall that we have chosen in the examples the same event horizon for the 4 solutions, and the same distances $r_1, r_2$, and $d$. Also, note that that when the unit is not specified, we are considering geometrized units. Namely, mass, charge and time have unit of length (m).
As we can see, for the scale of stellar black holes, there is no difference in the results down to the order of milliseconds. The shortest delay was for Bardeen solution, and the longest for Reissner-Nordstr\"o{}m. This is interesting because it shows that; even respecting the critical charge condition, $q_{BD}< 4/3\sqrt{3} m_3$, and having a slightly larger mass; the Bardeen solution has the delay smaller than Schwarzschild. This is due to the negative parts in the expression \eqref{eq:delayBD}, which are always attached to the magnetic charge. The results we got for the solutions of Reissner-Nordstr\"o{}m and ABG were the closest, which was  expected since their analytic  expressions \eqref{eq:ttoltalRN} and \eqref{eq:ttoltalABG} are identical.

If we now consider supermassive black holes, with the masses
\begin{equation}
    m_2 = m_3 = m_4 = 3.6 \times 10^9 \times M_{\odot}^{\prime},
\end{equation}
and event horizon
\begin{equation}
    r_+ = 1.00916 \times 10^{13}.
\end{equation}
Also, by imposing the same values used in the last section for the distances, that is
\begin{equation}
    \begin{aligned}
       & d = 2.7 \times 10^{13}, \\
       & r_1 = 2.0 \times 10^2d, \\
       & r_2 = 1.0 \times 10^2d.
    \end{aligned}
\end{equation}
We have the following results displayed in the table \ref{tabela2}. The event horizon for the 4 solutions, and the distances $r_1, r_2$, and $d$ are the same. Again, there is nothing special with these parameter choices, they are just values that satisfy the imposed conditions. Any other choices, whether they come from some real situation or not, lead to the same main result we find here: the difference between the Shapiro time for these solutions.  The Bardeen solution again showed the shortest delay, however, the biggest was for the ABG solution. The difference between these results is approximately  $58.7$ hours. All delays were in the order of days, this is because the expressions are proportional to the mass and we use masses billions of times greater than the mass of the Sun.  Backing to our motivating question, is it  possible to distinguish between different black hole solutions using the Shapiro time delay? We conclude that the solutions are distinguishable analytically, except Reissner-Nordstr\"o{}m and Ayón-Beato and García that are equal. Furthermore, they are also distinguishable taking into account an idealized numerical model.

\begin{table}[htpb]
\centering
\caption{Stellar black holes}
\label{tabela1}
\vspace{0.5cm}
\begin{tabular}{|c|c|c|c|c|}
\hline
\textbf{Solution} & \textbf{Mass $\left(M^{\prime}_{\odot} \right)$}   & \textbf{Charge ($m$)}   &  \textbf{Total Time (s)}    & \textbf{Delay Time (ms)}          \\ 
       \hline                       
 Schwarzschild &$6.0$ & $0$& $9.32$ & $2.55261$  \\ 
\hline    
Reissner-Nordstr\"o{}m & $6.1$ & $2299.1$  & $9.32$ & $2.59337$ \\
\hline
Bardeen &   $6.1$ &  $1874.62$ & $9.32$ & $2.23284$    \\
\hline    
ABG &   $6.1$ & $ 1454.32$ & $9.32$ & $2.59159$       \\
\hline    

\end{tabular}
\end{table}

\begin{table}[htpb]
\centering
\caption{Supermassive black holes}
\label{tabela2}
\vspace{0.5cm}
\begin{tabular}{|c|c|c|c|c|}
\hline
\textbf{Solution} & \textbf{Mass $\left(10^9 \times M^{\prime}_{\odot} \right)$}   & \textbf{Charge ($m$)}    & \textbf{Total Time (s)}    & \textbf{Delay Time (s)}          \\ 
       \hline                       
 Schwarzschild &$3.4$ & $0$ & $5.4825 \times 10^7$ & $826827$  \\ 
\hline    
Reissner-Nordstr\"o{}m & $3.6$ & $2.44758 \times 10^{12}$ & $5.48666 \times 10^7$ & $868527$ \\
\hline    
Bardeen &   $3.6$ &  $1.98887\times 10^{12}$  & $5.48713 \times 10^7$ & $661373$    \\
\hline    
ABG &   $3.6$ & $1.54885\times 10^{12}$ & $5.48708 \times 10^7$ & $872686$       \\
\hline    

\end{tabular}
\end{table}

\section{CONCLUSION \label{sec:six}}

In this paper we calculate the Shapiro time delay for four solutions in order to investigate the possibility of distinguishing between them. The motivation came from the idea present in some papers that regular solutions would be indistinguishable from standard solutions \cite{Paula:2020yfr,dePaula:2022kzz}. Of course, there are also mentions to the contrary \cite{Stuchlik:2019uvf,Lima:2021las}; the results here point in that direction. The solutions used were Schwarzschild, has only mass; Reissner-Nordstr\"o{}m (RN), has mass and electric charge; Bardeen, has mass and magnetic charge; and Ayón-Beato and García (ABG), has mass and electric charge. The Schwarzschild and RN black holes are singular, while the Bardeen and ABG black holes are regular. We point out here some of the main characteristics of these solutions. We also show how to get Shapiro time for any metric. This subject is  common in the literature, and is even covered in some textbooks because it is one of the tests of general relativity. However, as far as we know, this is the first time that Shapiro time has been calculated for Bardeen and ABG solutions.

We started, in section \ref{sec:one}, showing the metrics and some properties of the black hole solutions we were going to use.  In section \ref{sec:two},
we assume a idealized situation similar to that used in experiments used to test the general relativity \cite{Will:2014kxa}. 
Therefore, we seek to calculate the time it takes for a light signal to go from one planet to another when they are in a configuration called superior conjunction, that is, making a straight line passing through some more massive body in between (the Sun, some star, or a black hole). The calculated time is longer than if the massive body was not present, that is, there is a delay in the relativistic time relative to the time for flat space. In other words, we show how to calculate Shapiro time delay for general static and spherically symmetric metric.  The result depends on the parameters of the solution; in the case of Schwarzschild we have only the mass, the distance of closest approach to the central body $d$; and the distances from $d$ to the planets, labeled $r_1$ and $r_2$.

To compare the observational result with his theoretical prediction Shapiro used $d= R_{\odot}$ \cite{Inverno}, in the case of black holes the most intuitive would then use $d=r_+$. However, as we showed in the section \ref{sec:three}, photons are captured by these bodies at distances lesser or equal to the critical impact parameter, which is usually larger than the event horizon.  
 We calculate the critical impact parameters for general static and spherically symmetric metric, and in our further numerical examples we choose $d$ to be greater than the $b_c$ of the solutions involved.

 In section \ref{sec:four} we  compare the Shapiro time of Schwarzschild and RN solutions. The analytic results for the time delay were obtained by considering first-order expansions of the mass, Schwarzschild, and the gravitation constant $G$, RN. The expression for RN solution return the Schwarzschild expression when the charge is zero, and we noticed that result showed symmetry at charge $q \rightarrow - q$, and that it acts in a way that decreases the retardation. There could be some configuration of charge and mass in which the relativistic time is smaller than the flat time, however, we removed this possibility of having an advance by always considering charges smaller than the critical one. 
 We did two types of numerical examples, one considering masses of the order of the mass of the Sun, stellar black holes, and another considering masses billions of times greater than the mass of the Sun, supermassive black holes. We chose the values of the masses respecting the condition \eqref{eq:carga1}, and then we found the value of the charge of the RN solution imposing that both bodies had the same event horizon. With the chosen value for the masses we calculate the critical impact parameter for both solutions and choose $d$, $r_1$ and $r_2$.
 We find a difference in the delay time from the order of $10^{-5}s$ for the first example. Since the expressions \eqref{eq:delaySC} and \eqref{eq:delayRN} are proportional to mass, the second example generated a difference in delay time of the order of hours.

 In section \ref{sec:five} we compare RN, Bardeen and ABG solutions. The analytic results for the Shapiro time delay were obtained by considering first-order expansions of the mass, Bardeen, and the gravitation constant $G$, RN and ABG. The expressions for all charged  solutions return the Schwarzschild expression when the charge is zero. Also, they possess symmetry at charge $q \rightarrow - q$, and  it acts in a way that decreases the retardation. This is an effect opposite to that of mass. For the numerical analysis
 we adopt a similar procedure as in the previous section, except that now the masses of the black holes are equal. With that said, and choosing the same event horizon value found earlier, we calculated the charges. The solutions of RN and ABG, which have equal expressions for travel time, had different electrical charges. Still, the times found for these solutions were the closest. The time delay considering stellar black hole, for example, can only be deferred on the order of micro seconds ($10^{-6}$s). Bardeen's solution generated the least time delay in the two cases considered, this reveals that the magnetic charge has an opposite effect to the mass, that is, it decreases the travel time. In fact, this is also true for the electric charges in \eqref{eq:delayRN} and \eqref{eq:delayABG}, but, the numerical values we chose and the critical charge boundary, caused the mass term to dominate these expressions. 
 We compiled all the information found in the tables  \ref{tabela1} and \ref{tabela2}.  Despite using a idealized numerical model, we believe that the Shapiro time could be used to distinguish between different types of black hole solutions.

In a future work, we hope to improve our results by using larger approximation orders,  and adding the correction for velocity terms $v/c$\cite{He:2016cya}. This could lead, for example, to an analytical difference in the results of Reissner-Nordstr\"o{}m and  Ayón-Beato and García.  Another perspective is to check what effect the rotation has on Shapiro time delay.  There are already some regular solutions that  include rotation \cite{Bambi:2013ufa,Toshmatov:2014nya,Neves:2014aba}; and a recent work \cite{Kumar:2020yem}  showed, by shadow analysis,  that Bardeen-type, Hayward-type \cite{Hayward:2005gi}, and Culetu-type \cite{Culetu:2014lca} solutions with rotation are indistinguishable from Kerr solution.

 
\vspace{1cm}

{\bf Acknowledgements}: M. E. R.  thanks Conselho Nacional de Desenvolvimento Cient\'ifico e Tecnol\'ogico - CNPq, Brazil, for partial financial support. This study was financed in part by the Coordena\c{c}\~{a}o de Aperfei\c{c}oamento de Pessoal de N\'{i}vel Superior - Brasil (CAPES) - Finance Code 001.


\end{document}